\documentstyle[12pt,preprint,aps]{revtex}
\newcommand{\lag}[1]{{\cal L}_g^{#1}}
\newcommand{\Lag}{{\cal L}_g}
\newcommand{\riemann}[4]{ R^{#1}_{{#2}{#3}{#4}} }
\newcommand{\ricci}[2]{R_{{#1}{#2}}}
\newcommand{\affine}[3]{\Gamma^{#1}_{{#2}{#3}}}
\newcommand{\fk}[8]{\partial{#1}h{#2}{#3}\partial{#4}h{#5}{#6}h{#7}{#8}}
\begin{document}
\draft
\vskip 2.5cm
\title{\bf
The simplest form of the $\kappa$ and $\kappa^2$ order\\
graviton self-interaction Lagrangian density \\
in the weak field approximation
}
\vskip 3.5cm
\author{
\renewcommand\thefootnote{\alph{footnote}}
\setcounter{footnote}{0}
Jungil Lee$^{1,}$\footnote{jungil@phyy.snu.ac.kr},
J. S. Shim$^{2,}$\footnote{shim@phyy.snu.ac.kr},
and
\renewcommand\thefootnote{\alph{footnote}}
\setcounter{footnote}{2}
H. S. Song$^{1,}$\footnote{hssong@phyy.snu.ac.kr}
}
\address{$^{1}$ Center for Theoretical Physics and Department of Physics, \\
                Seoul National University, Seoul 151-742, Korea}
\address{$^{2}$ Department of Physics, Hanyang University, Seoul , Korea}
\vskip 2.5cm
\maketitle
\begin{abstract}
We derive the simplest form of the $\kappa(\kappa^2)$  order graviton
self-interaction Lagrangian density
$\lag{1}(\lag{2})$
in the weak field approximation.
With the divergenceless condition, de Donder gauge and some combinatoric
techniques, we derive the mathematically independent basis expressions for
$\lag{1}(\lag{2})$
composed of 8(24) terms.
By using the basis expressions
$\lag{1}(\lag{2})$
is reduced to 6(18) terms.
\end{abstract}
\renewcommand{\thepage}{}
\newpage
\renewcommand{\thepage}{\arabic{page}}\setcounter{page}{1}
\begin{center}{\bf 1.~Introduction}\end{center}

Whenever one studies quantum gravity processes
in the weak field approximation procedure\cite{Gup}, where
the metric perturbation $h_{\mu\nu}$ in $g_{\mu\nu}$
from the flat space metric
$\eta_{\mu\nu}$ is identified with the gravitational field,
one always encounters the difficulties in dealing with
the complexity of the algebraic work
which stems from the complicated nature of the graviton.
A large number of indices appear in the interaction Lagrangian
since a graviton has a pair of indices and it can even experience
self-interaction.
Furthermore, one should carry out numerous
permutation and symmetrization
operations on each index-pair and over all momentum-index triplets
of the gravitons in the interaction Lagrangian
to get the correct vertices. This is due to
the symmetry  property of the index pair of $h_{\mu\nu}$
and the bosonic property of the graviton, respectively.
In 1967, DeWitt\cite{DeW} derived the graviton self-interaction
Lagrangian up to 3- and 4- graviton vertices. This Lagrangian has been
believed to be composed of the smallest terms but not been proved.
Using this Lagrangian, cross-sections for the processes involving
3- or 4- graviton vertices have been evaluated\cite{Ber,Vor}  without
further simplifying the Lagrangian
except for the marginal correction of the sign of a term\cite{Ber}
in the 4- graviton vertex function. As shown in section 2,
3- and 4- graviton vertices can be obtained from $\lag{1}$ and $\lag{2}$.

In this paper, we present the basis expressions for $\lag{1}$ and $\lag{2}$
and the simplest form of $\lag{1}$ and $\lag{2}$ in terms of the basis
expressions.
$\lag{1}$ is simplified from 11 to 6 in the number of terms.
For the case of  $\lag{2}$, 28 terms are reduced into only 18 terms.
Though the numbers of terms are still rather large,
they cannot be simplified any more
because they are expressed by the basis expressions.
Using our new $\lag{1}$ and $\lag{2}$,
we obtained the same differential cross-sections for the
graviton-scalar($gs\rightarrow gs$)\cite{Ber},
graviton-fermion($gf\rightarrow gf$)\cite{Vor},
graviton-massive vector boson($gW\rightarrow gW$)\cite{Ber},
graviton-photon($g\gamma\rightarrow g\gamma$)\cite{Ber} and
graviton-graviton($gg\rightarrow gg$)\cite{Ber}
elastic scatterings with those of the corresponding references.
In this paper, we review the formal procedure to obtain the
graviton self-interaction Lagrangian in the weak field approximation
in section 2.
To deal with the combinatoric properties of the typical terms
in the Lagrangian conveniently, we introduce our index convention which
is also helpful in computer-aided calculations in section 3.
Sections 4 and 5 are devoted to explain
our strategy for getting the simplest expressions.
In these sections, the scheme of constructing basis expressions for $\lag{1}$
and $\lag{2}$ and also the explicit forms of bases
as well as the simplest forms of
$\lag{1}$ and $\lag{2}$ are given in these sections.
Conclusion is made in section 6.
\begin{center}{\bf 2.~Analysis of the problem}\end{center}

In the classical theory of general relativity\cite{Weinberg}, the action for
pure gravity is given by
\begin{equation}
I_g=2 \kappa^{-2} \int \mbox{d}^4 x\sqrt{-g(x)}R(x),
\end{equation}
where $\kappa=\sqrt{32\pi G_N}$ with the Newtonian constant $G_N$, $R(x)$ the
scalar curvature,
and $g(x)=\det g_{\mu\nu}$, i.e., the graviton self-interaction
Lagrangian density is given by
\begin{equation}
\Lag=2\kappa^{-2}\sqrt{-g}R.
\end{equation}
In the weak field approximation, one can write
the metric tensor $g_{\mu\nu}$ as the sum of the flat space metric
$\eta_{\mu\nu}$ and a metric perturbation $h_{\mu\nu}$ as
\begin{equation}
g_{\mu\nu}=\eta_{\mu\nu}+\kappa h_{\mu\nu}.
\end{equation}
Here $h_{\mu\nu}$ can be identified with the gravitational field by
giving [mass]$^{-1}$ dimension to the parameter $\kappa$.
The curvature scalar $R$ is defined by
\begin{eqnarray}
R&=&g^{\mu\nu}\ricci{\mu}{\nu}\nonumber\\
 &=&g^{\mu\nu}\riemann{\lambda}{\mu}{\lambda}{\nu},
\end{eqnarray}
where $R_{\mu\nu}$ is the Ricci tensor and
the Riemann-Christoffel curvature tensor
$\riemann{\lambda}{\mu}{\kappa}{\nu}$
is defined by
\begin{equation}
\riemann{\lambda}{\mu}{\kappa}{\nu}
=\partial_{\nu}\affine{\lambda}{\mu}{\kappa}
-\partial_{\kappa}\affine{\lambda}{\mu}{\nu}
+\affine{\alpha}{\mu}{\kappa}\affine{\lambda}{\alpha}{\nu   }
-\affine{\alpha}{\mu}{\nu   }\affine{\lambda}{\alpha}{\kappa},
\end{equation}
and the affine connection $\affine{\lambda}{\mu}{\nu}$ is defined by
\begin{equation}
\affine{\lambda}{\mu}{\nu}=
\frac{1}{2}g^{\lambda\alpha}
(
 \partial_\mu    g_{\alpha\nu}
+\partial_\nu    g_{\alpha\mu}
-\partial_\alpha g_{\mu   \nu}
)
\end{equation}
By using the identity $g^{\mu\lambda}g_{\lambda\nu}=\delta^\mu_\nu$,
we get the $\kappa$ expansion of $g^{\mu\nu}$,
\begin{equation}
g^{\mu\nu}=
\eta^{\mu\nu}
-\kappa h^{\mu\nu}
+\kappa^2 h^{\mu\alpha} h_\alpha^\nu
-\kappa^3 h^{\mu\alpha} h_\alpha^\beta h_\beta^\nu+...
\end{equation}
Expanding $g$ and $\sqrt{-g}$ up to $\kappa^3$ order , we get
\begin{eqnarray}
g&=&
-1
-\kappa h
-\frac{1}{2}\kappa^2(h^2-h^{\mu\nu} h_{\mu\nu})
-\frac{1}{6}\kappa^3(h^3
-3 h h^{\nu\sigma} h_{\nu\sigma}
+2 h^{\mu\nu} h_\nu^\sigma h_{\sigma\mu}),\\
\sqrt{-g}&=&
1
+\frac{1}{2}\kappa h
+\frac{1}{8}\kappa^2(h^2-2h^{\mu\nu} h_{\mu\nu})
+\frac{1}{48}\kappa^3
(  h^3
-6 h h^{\nu\sigma} h_{\nu\sigma}
+8 h^{\mu\nu} h_\nu^\sigma h_{\sigma\mu}),
\end{eqnarray}
with the definition $h=h^\mu_\mu$.
Then we can obtain the leading term of $\Lag$ as
\begin{equation}
\Lag=2\kappa^{-2}
[\kappa^2(
-\frac{1}{8}\partial_\mu h \partial^\mu h
+\frac{1}{4}\partial_\mu h^\alpha_\beta\partial^\mu h_\alpha^\beta)
+O(\kappa^3)],
\end{equation}
where we use the harmonic(de Donder) gauge,
\begin{equation}
\partial_\mu h^\mu_\alpha=\frac{1}{2}\partial_\alpha h,
\end{equation}
and omit total derivatives.
Then $\kappa$ expansion of Lagrangian density is given by
\begin{eqnarray}
\Lag&=&\frac{2}{\kappa^2}\sqrt{-g}R\nonumber\\
    &=&\lag{0}+\kappa\lag{1}+\kappa^2\lag{2} +...
\end{eqnarray}
The leading term $\lag{0}$ is obviously of the simplest form as
\begin{equation}
\lag{0}=
-\frac{1}{4}\partial_\mu h\partial^\mu h
+\frac{1}{2}\partial_\mu h^\alpha_\beta\partial^\mu h_\alpha^\beta.
\end{equation}
$\lag{1}$ and $\lag{2}$ can be arranged into simpler forms
by using some conditions explained below.
According to Refs.\cite{DeW,Ber}, they are reduced into 11 and 28 terms,
respectively.
But we have obtained still more simplified forms of
$\lag{1}$ and $\lag{2}$ and proved that
they cannot be reduced any more.

The framework of our proof is as follows:
Let ${\cal E}^i(i=1,2)$ be the space composed of
any mathematically possible expressions
which has the typical form of the terms inside $\lag{i}$.
(in this paper any upper indices $i, j, k, l$ are related only to
the order of $\kappa$.)
To get the simplest form of $\lag{i}$, we should find out the basis
$B^i$ for ${\cal E}^i$.
We start our search for $B^i$ from a set $C^i$
which contains all the possible terms
inside any expression in ${\cal E}^i$.
Of course $C^i$ spans ${\cal E}^i$ and has
all the possible permutaions of indices for the typical terms.
Though $C^i$ has a large number of elements,
it is both countable and finite.
Without any condition and neglecting the multiplication commutativity,
the numbers of elements of $C^i$'s are
\begin{eqnarray}
C^1&:&\frac{8!}{(2!)^4\cdot4!},\nonumber\\
C^2&:&\frac{10!}{(2!)^5\cdot5!}.
\end{eqnarray}
With the symmetry property of $h_{\mu\nu}$
and de Donder gauge condition,
we can reject a lot of duplicate elements from  $C^i$.
If $C^i$ has $n^i$ elements in this step,
we can express $C^i$ as
\begin{equation}
C^i=\{c^i_1,c^i_2,c^i_3,...,c^i_{n^i}\}.
\end{equation}
We can obtain the useful relations
from the fact that the total derivative
of the Lagrangian can be neglected.
Since any $c^i_j$ has derivative factors,
we can rewrite it by partial integration
neglecting the total derivatives.
In general the result of this partial integration is also an element of
${\cal E}^i$. Therefore, we can express each $c^i_j$ as an linear
combination of $c^i_k$'s$(k=1,2,3,...,n^i)$ as
\begin{equation}
c^i_j=\sum_{k=1}^{n^i} d^i_{jk} c^i_k,
\end{equation}
where $d^i_{jk}$'s are pure numbers.
If we solve these $n^i$-simultaneous equations, we can find the
basis. Not all the equations are independent and
some $c^i_j$'s are rewritten by itself under partial integration
as
\begin{equation}
c^i_l=c^i_l.
\end{equation}
If there are $m^i$ non-trivial equations in the $n^i$-simultaneous
equations, there exist $n^i-m^i$ free variables in the simultaneous
equations. Therefore we can take any $n^i-m^i$ elements from
$C^i$ to make a basis set $B^i$.
It is a common orthogonalization procedure.
Getting the basis expressions, we can obtain the simplest form
of the Lagrangians by expressing the original $\lag{1}$ and $\lag{2}$
in terms of these basis expressions.
Let us find out them step by step in the following sections.
\begin{center}{\bf 3.~Convention Summary}\end{center}

First of all, we introduce our new convention for expressions
appearing in $\lag{1}$ and $\lag{2}$.
Through partial integration and omitting total derivative, every term of
$\lag{1}$ and $\lag{2}$ can be transformed into the
following typical forms,
\begin{eqnarray}
\lag{1}&:&\partial_{?}h_{??}\partial_{?}h_{??}h_{??}\nonumber\\
\lag{2}&:&\partial_{?}h_{??}\partial_{?}h_{??}h_{??}h_{??},
\end{eqnarray}
where ?'s are the unknown indices. There are 4 and 6 pairs of indices
in the typical terms of $\lag{1}$ and $\lag{2}$, respectively.
Here we neglect the difference between the upper and lower indices
because every pair of indices is composed of 1 upper index and
1 lower one - contraction is always carried out between them.
To show the combinatoric characteristics of the indices,
we use a number in place of a Greek character as an index.
Here is an explicit example of our convention.
\begin{eqnarray}
                 h_{\mu\nu} &\rightarrow& (12)\nonumber\\
\partial_{\alpha}h_{\mu\nu} &\rightarrow& (3(12))
\end{eqnarray}
Distinct numbers should be given to the different indices.
Innermost parenthesis means $h_{??}$.
If there is a double parenthesis,
outer one means derivative. Every number in the parenthesis
means the correspondent index.
For brevity, double derivative expression like (1(2(34)))
is written as (12(34)).
This convention is very convenient to read and write in the
computer-aided calculations.
Especially, it shows the combinatoric characteristics of each
expressions much more clearly.
\newpage
\begin{center}{\bf 4.~$\lag{1}$ reduction}\end{center}

For the case of $\lag{1}$, the general form is
$(a(l_1l_2))(b(m_1m_2))(n_1n_2)$.
After listing all possible combinations of
indices made of 4 pairs of numbers, we can find the projection
among the expressions by using partial integration and
neglecting total derivatives as
\begin{eqnarray}
  &(&a(l_1l_2))(b(m_1m_2))(n_1n_2)\nonumber\\
=-&(&l_1l_2)(ab(m_1m_2))(n_1n_2)-(l_1l_2)(b(m_1m_2))(a(n_1n_2))\nonumber\\
=+&(b(&l_1l_2))(a(m_1m_2))(n_1n_2)-(l_1l_2)(b(m_1m_2))(a(n_1n_2))\nonumber\\
 +&(&l_1l_2)(a(m_1m_2))(b(n_1n_2)).
\end{eqnarray}
Throughout this procedure we extensively use the de Donder gauge condition
to get the factor such as (11) or (1(22)) maximally.
Rejecting the dependent expressions one by one, we get the basis
expressions which can only be transformed into itself.
They are given by
\begin{eqnarray}
f^1_1&=&(1(22))(1(33))(44)=
\fk{^\mu}{^\nu}{_\nu}{_\mu}{^\alpha}{_\alpha}{^\beta}{_\beta}\nonumber\\
f^1_2&=&(1(22))(4(33))(14)=
\fk{^\mu}{^\nu}{_\nu}{^\beta}{^\alpha}{_\alpha}{_\mu}{_\beta}\nonumber\\
f^1_3&=&(1(22))(1(34))(34)=
\fk{^\mu}{^\nu}{_\nu}{_\mu}{^\alpha}{^\beta}{_\alpha}{_\beta}\nonumber\\
f^1_4&=&(1(22))(3(14))(34)=
\fk{^\mu}{^\nu}{_\nu}{^\alpha}{_\mu}{^\beta}{_\alpha}{_\beta}\nonumber\\
f^1_5&=&(1(23))(1(23))(44)=
\fk{^\mu}{^\nu}{^\alpha}{_\mu}{_\nu}{_\alpha}{^\beta}{_\beta}\nonumber\\
f^1_6&=&(1(23))(4(23))(14)=
\fk{^\mu}{^\nu}{^\alpha}{^\beta}{_\nu}{_\alpha}{_\mu}{_\beta}\nonumber\\
f^1_7&=&(1(23))(2(34))(14)=
\fk{^\mu}{^\nu}{^\alpha}{_\nu}{_\alpha}{^\beta}{_\mu}{_\beta}\nonumber\\
f^1_8&=&(1(23))(1(24))(34)=
\fk{^\mu}{^\nu}{^\alpha}{_\mu}{_\nu}{^\beta}{_\alpha}{_\beta}.
\end{eqnarray}
The basis is made up of 8 expressions. This number is smaller than that(11)
of the old $\lag{1}$\cite{Ber}.
If one expresses $\lag{1}$ in terms of this basis,
it is composed of even smaller number of terms
, i.e., 6 terms as
\begin{equation}
\lag{1}=
-\frac{1}{8}f^1_1
+\frac{1}{2}f^1_3
+\frac{1}{4}f^1_5
-\frac{1}{2}f^1_6
+           f^1_7
-           f^1_8.
\end{equation}
\begin{center}{\bf 5.~$\lag{2}$ reduction}\end{center}
The method of $\lag{2}$ reduction is basically the same
as that of $\lag{1}$.
The general form of a term in $\lag{2}$ is of the form,
$(a(l_1l_2))(b(m_1m_2))(n_1n_2)(r_1r_2)$,
which has one more pair of indices with one more $h_{??}$.
There are the same number of derivative factors which make
the transformation rule alike.
We can find the projection of a typical term to the others
by using partial integration and neglecting total derivatives as
\begin{eqnarray}
  &(&a(l_1l_2))(b(m_1m_2))  (n_1n_2)  (r_1r_2)\nonumber\\
=-&(&  l_1l_2)(ab(m_1m_2))  (n_1n_2)  (r_1r_2)\nonumber\\
 -&(&  l_1l_2) (b(m_1m_2))(a(n_1n_2)) (r_1r_2)\nonumber\\
 -&(&  l_1l_2) (b(m_1m_2))  (n_1n_2)(a(r_1r_2))\nonumber\\
=+&(&b(l_1l_2))(a(m_1m_2))  (n_1n_2)  (r_1r_2)\nonumber\\
 +&(&  l_1l_2) (a(m_1m_2))(b(n_1n_2)) (r_1r_2)\nonumber\\
 +&(&  l_1l_2) (a(m_1m_2))  (n_1n_2)(b(r_1r_2))\nonumber\\
 -&(&  l_1l_2) (b(m_1m_2))(a(n_1n_2)) (r_1r_2)\nonumber\\
 -&(&  l_1l_2) (b(m_1m_2))  (n_1n_2)(a(r_1r_2)).
\end{eqnarray}
In this case, we can find the projections of a typical term to
5 other terms.
Rejecting the dependent expressions, we get the basis
expressions,
\begin{eqnarray}
f^2_1&=&(1(22))(1(33))(44)(55)=
\fk{^\mu}{^\nu}{_\nu}{_\mu}{^\alpha}{_\alpha}{^\beta}{_\beta}
h{^\sigma}{_\sigma}\nonumber\\
f^2_2&=&(1(22))(1(33))(45)(45)=
\fk{^\mu}{^\nu}{_\nu}{_\mu}{^\alpha}{_\alpha}{^\beta}{^\sigma}
h{_\beta}{_\sigma}\nonumber\\
f^2_3&=&(1(22))(3(44))(13)(55)=
\fk{^\mu}{^\nu}{_\nu}{^\alpha}{^\beta}{_\beta}{_\mu}{_\alpha}
h{^\sigma}{_\sigma}\nonumber\\
f^2_4&=&(1(22))(3(44))(15)(35)=
\fk{^\mu}{^\nu}{_\nu}{^\alpha}{^\beta}{_\beta}{_\mu}{^\sigma}
h{_\alpha}{_\sigma}\nonumber\\
f^2_5&=&(1(22))(1(34))(34)(55)=
\fk{^\mu}{^\nu}{_\nu}{_\mu}{^\alpha}{^\beta}{_\alpha}{_\beta}
h{^\sigma}{_\sigma}\nonumber\\
f^2_6&=&(1(22))(1(34))(35)(45)=
\fk{^\mu}{^\nu}{_\nu}{_\mu}{^\alpha}{^\beta}{_\alpha}{^\sigma}
h{_\beta}{_\sigma}\nonumber\\
f^2_7&=&(1(22))(3(14))(34)(55)=
\fk{^\mu}{^\nu}{_\nu}{^\alpha}{_\mu}{^\beta}{_\alpha}{_\beta}
h{^\sigma}{_\sigma}\nonumber\\
f^2_8&=&(1(22))(3(14))(35)(45)=
\fk{^\mu}{^\nu}{_\nu}{^\alpha}{_\mu}{^\beta}{_\alpha}{^\sigma}
h{_\beta}{_\sigma}\nonumber\\
f^2_9&=&(1(22))(3(45))(45)(13)=
\fk{^\mu}{^\nu}{_\nu}{^\alpha}{^\beta}{^\sigma}{_\beta}{_\sigma}
h{_\mu}{_\alpha}\nonumber\\
f^2_{10}&=&(1(22))(3(45))(41)(53)=
\fk{^\mu}{^\nu}{_\nu}{^\alpha}{^\beta}{^\sigma}{_\beta}{_\mu}
h{_\sigma}{_\alpha}\nonumber\\
f^2_{11}&=&(1(23))(2(13))(45)(45)=
\fk{^\mu}{^\nu}{^\alpha}{_\nu}{_\mu}{_\alpha}{^\beta}{^\sigma}
h{_\beta}{_\sigma}\nonumber\\
f^2_{12}&=&(1(23))(1(23))(44)(55)=
\fk{^\mu}{^\nu}{^\alpha}{_\mu}{_\nu}{_\alpha}{^\beta}{_\beta}
h{^\sigma}{_\sigma}\nonumber\\
f^2_{13}&=&(1(23))(1(23))(45)(45)=
\fk{^\mu}{^\nu}{^\alpha}{_\mu}{_\nu}{_\alpha}{^\beta}{^\sigma}
h{_\beta}{_\sigma}\nonumber\\
f^2_{14}&=&(1(23))(1(24))(34)(55)=
\fk{^\mu}{^\nu}{^\alpha}{_\mu}{_\nu}{^\beta}{_\alpha}{_\beta}
h{^\sigma}{_\sigma}\nonumber\\
f^2_{15}&=&(1(23))(1(24))(35)(45)=
\fk{^\mu}{^\nu}{^\alpha}{_\mu}{_\nu}{^\beta}{_\alpha}{^\sigma}
h{_\beta}{_\sigma}\nonumber\\
f^2_{16}&=&(1(23))(1(45))(23)(45)=
\fk{^\mu}{^\nu}{^\alpha}{_\mu}{^\beta}{^\sigma}{_\nu}{_\alpha}
h{_\beta}{_\sigma}\nonumber\\
f^2_{17}&=&(1(23))(1(45))(24)(35)=
\fk{^\mu}{^\nu}{^\alpha}{_\mu}{^\beta}{^\sigma}{_\nu}{_\beta}
h{_\alpha}{_\sigma}\nonumber\\
f^2_{18}&=&(1(23))(4(23))(14)(55)=
\fk{^\mu}{^\nu}{^\alpha}{^\beta}{_\nu}{_\alpha}{_\mu}{_\beta}
h{^\sigma}{_\sigma}\nonumber\\
f^2_{19}&=&(1(23))(4(23))(15)(45)=
\fk{^\mu}{^\nu}{^\alpha}{^\beta}{_\nu}{_\alpha}{_\mu}{^\sigma}
h{_\beta}{_\sigma}\nonumber\\
f^2_{20}&=&(1(23))(4(25))(13)(45)=
\fk{^\mu}{^\nu}{^\alpha}{^\beta}{_\nu}{^\sigma}{_\mu}{_\alpha}
h{_\beta}{_\sigma}\nonumber\\
f^2_{21}&=&(1(23))(4(25))(14)(35)=
\fk{^\mu}{^\nu}{^\alpha}{^\beta}{_\nu}{^\sigma}{_\mu}{_\beta}
h{_\alpha}{_\sigma}\nonumber\\
f^2_{22}&=&(1(23))(2(34))(15)(45)=
\fk{^\mu}{^\nu}{^\alpha}{_\nu}{_\alpha}{^\beta}{_\mu}{^\sigma}
h{_\beta}{_\sigma}\nonumber\\
f^2_{23}&=&(1(23))(2(34))(14)(55)=
\fk{^\mu}{^\nu}{^\alpha}{_\nu}{_\alpha}{^\beta}{_\mu}{_\beta}
h{^\sigma}{_\sigma}\nonumber\\
f^2_{24}&=&(1(23))(2(45))(14)(35)=
\fk{^\mu}{^\nu}{^\alpha}{_\nu}{^\beta}{^\sigma}{_\mu}{_\beta}
h{_\alpha}{_\sigma}.
\end{eqnarray}
The basis is made up of 24 expressions. This number is smaller than that(28)
of the old $\lag{2}$. If one expresses $\lag{2}$ in terms of this basis,
it is composed of even smaller number of terms
, i.e., 18 terms as
\begin{eqnarray}
\lag{2}&=&
-\frac{1}{32}f^2_1
+\frac{1}{ 8}f^2_2
+\frac{1}{ 4}f^2_5
-\frac{1}{ 2}f^2_6
+\frac{1}{ 2}f^2_8
-\frac{1}{ 4}f^2_{11}
+\frac{1}{16}f^2_{12}
-\frac{1}{ 8}f^2_{13}
-\frac{1}{ 2}f^2_{14}\nonumber\\
&&+            f^2_{15}
-\frac{1}{ 2}f^2_{16}
+\frac{1}{ 2}f^2_{17}
-\frac{1}{ 4}f^2_{18}
+\frac{1}{ 2}f^2_{19}
-            f^2_{20}
+            f^2_{21}
-2           f^2_{22}
+\frac{1}{ 2}f^2_{23}.
\end{eqnarray}
This reduction in the number of terms in $\lag{1}$ and $\lag{2}$
can not be done any further.
The simplest form shows her face after 27 year's latency.
Though the idea we employed to get the results is somewhat simple,
the actual procedure are hard to be carried out properly
if one sticks to manual work only.
Throughout our calculation,
we extensively used a symbolic manipulation package
{\it Mathematica}\cite{Math}
which enabled us to execute the powerful pattern matching programs.
\begin{center}{\bf 6.~Conclusion}\end{center}

The simplified forms
of the graviton self-interaction Lagrangian densities are derived
to the $\kappa$ and $\kappa^2$ order
and they are proved to be the simplest.
The key point is that we first construct the basis expressions
for the $\lag{1}$ and $\lag{2}$ and express them by this basis.
We presented these basis sets explicitly and
described not only the practical way to get the basis
but also the formal proof.
We also checked our results by calculating the observables of
several real processes.
To check the 3-graviton vertex, we evaluated the differential
cross-sections for the processes,
graviton-scalar($gs\rightarrow gs$)\cite{Ber},
graviton-fermion($gf\rightarrow gf$)\cite{Vor},
graviton-massive vector boson($gW\rightarrow gW$)\cite{Ber},
graviton-photon($g\gamma\rightarrow g\gamma$)\cite{Ber},
by using our simplest $\lag{1}$.
The differential cross-section of the process
graviton-graviton($gg\rightarrow gg$)\cite{Ber}
is also evaluated in the same way
to check the 3-and 4-graviton vertices.
We obtained the same results as those of the previous
graviton self-interaction Lagrangian.
With the simplified graviton self-interaction
Lagrangian density,
we can save much time and avoid errors in dealing with the numerous
indices.
And our new method to find the basis
expressions for the $\lag{1}$ and $\lag{2}$ can be applied to
any order of $\kappa$ with slight modification.
We presented a successful application of the powerful pattern matching
abilities of the symbolic manipulation package $Mathematica$
to  dealing with the physical problems involving complicated index relations.
\begin{center}{\bf Acknowledgements}\end{center}

The work was supported in part by the Korea Science and Engineering
Foundation  through the SRC program and
in part by the Korean Ministry of Education.

\end{document}